# Planar Dipole Antenna Design At 1800MHz Band Using Different Feeding Methods For GSM Application

Waleed Ahmed AL Garidi, Norsuzlin Bt Mohad Sahar, Rozita Teymourzadeh, CEng. *Member IEEE/IET*
Faculty of Engineering, Technology & Built Environment
UCSI University, 56000, Malaysia
Email: rozita@ucsi.edu.my

*Abstract-* **This research work focuses on the design and simulation of planar dipole antenna for 1800MHZ Band for Global System Mobile GSM application using Computer Software Technology CST studio software. The antenna is structured on fire resistance FR4 substrate with relative constant of 4.3 S/m. Two types of feeding configuration are designed to feed the antenna in order to match 50 Ω transmission lines which are via-hole integrated balun and quarter wavelength open stub. The via-hole is capable to provide maximum return loss of -25dB, bandwidth of 18.4% and the voltage standing wave ratio (VSWR) of 1.116 V at optimum dimension of length 59mm and width 4mm; the bandwidth is improved 25% to 30% by extending the width of the antenna 8 mm to 10 mm followed by deterioration of return loss value to -15dB. While the open stub at length of 67 mm, width of 6 mm and height 1.6mm will provide max return loss of -47.88dB and bandwidth of 17% with VSWR 1.008 << 2. Way that the antenna substrate has influenced the performance of the antenna. The lower relative constant will result the higher return lows, narrower bandwidth and better radiation pattern in trade-off the resonant length Via-hole and then the quarter wave open stub are most convenient for practical implementation.**

I. INTRODUCTION

Antenna is playing significant role in the wireless communication systems. Radio frequency and Microwave has direct effect on antenna for trance-receiving the signals and antenna will transmit it through the electro-magnetite wave in free space [1]. In order to design antenna in the bandwidth of 900MHz – 1800 MHz in the industry of cellular network such as GSM network, the IEEE Standard [1] (Std 145-1983) is approached accordingly.

The trends of the mobile phone technology has been dramatically decreased the weight and size of the communication equipment. Microstrip dipole antennas consist of a thin sheet of low loss insulating called the dielectric substrate and it is completely covered with a metal on one side called the ground plane.

From the other side is partly metalized where the circuit of antenna is printed [2]. In particular, planar dipole-type exhibits many attractive features, such as a simple structure, inexpensive easy integration with monolithic microwave integrated circuits (MMIC), low-profile, comfortable to planar and non-planar surfaces. Therefore, it works best on air and portable application. Micro-strip dipole moment is attractive because they basically own large feature like simple analysis and manufacturing and its attractive radiation pattern particularly low between-polarized rays.

II. ANTENNA GEOMETRY DESIGN

The desired frequency that the dipole antenna is designed to operate on is 1800MHz GSM band. The substrate material has been used is FR4 with dielectric constant of 4.3 thickness of 1.6mm.

The length of the antenna is the parameter that controls the resonant frequency. As result, the length is treated as half wave dipole, which is given by the following formula:

$$L = \frac{\lambda}{2} \qquad (1)$$

$$\lambda = \frac{C}{F} \qquad (2)$$

Where L is dipole length, λ is the wavelength, C is the speed of light in free space and F is the frequency of operation [3].

Table 1 shows the range of the dimensions parameters and table 2 presents the calculated values.

TABLE I
DESIGN RECOMMENDATION AND RESTRICTION [2]

| Parameters Symbol | Recommendation | Restriction |
|---|---|---|
| Width (W) | $0.05\lambda \leq W \leq 0.1\lambda$ | $0.05 \leq \frac{w}{h} \leq 20$ |
| Length (L) | $L \geq 0.48\lambda$ | $T/W \leq 0.5$ |
| Height (H) | $h \leq 0.02\lambda$ | $T/h \leq 0.5$ |
| Thickness (T) | $T \triangleleft \lambda$ | $\varepsilon r \geq 1$ |

TABLE II
CALCAULATED VALUES OF ANTENNA PARAMETERS

| Material Type | $\varepsilon_r$ | $\varepsilon_e$ | $\lambda_{d}$ mm | L-mm | w-mm | h-mm |
|---|---|---|---|---|---|---|
| FR4 | 4.3 | 2.65 | 104 | 52 | 6.25 | 2.08 |
| Arlon AD300 | 3 | 2 | 118 | 59 | 7.08 | 2.36 |
| Rogers RT5880 | 2.2 | 1.6 | 132 | 65.88 | 7.92 | 2.64 |

However, the speed of the signal when it propagates in free space varies as it propagates in medium and that is because of the effective dielectric constant of the microstrip substrate associated with fringing fields [4]. Therefore, the above equations have been modified to the following:

$$\lambda = \frac{c}{f\sqrt{\varepsilon}} \quad (3)$$

$$L = \frac{3\lambda}{4} \quad (4)$$

$$L = \frac{2\lambda}{3} \quad (5)$$

In most practical design, the wavelength of the printed dipole is treated as approximately in medium with equivalent to the average of that in free space, which is considered as alternative method for better determination of the length. Besides, the antenna parameters differ widely for different types of materials [5]. The feeding method is designed using integrated microstrip balun.

Planar antennas are integrated with other microwave circuitry. Therefore, the feeding techniques need to be designed and analyzed properly to reach between a good antenna performance and efficient circuit design. The word balun is a narrowing for "balanced to unbalanced" [5]. Table 3 has the practical parameters of the antenna being designed. At 1800 MHz band the antenna wavelength is 89.9 mm given by (3), the corresponding resonant length is 67mm for dipole with open stub and 60mm for dipole with via-hole by (4) and (5). Fig. 1 shows the open stub feeding while Fig. 2 illustrates the dipole antenna with via-hole feed.

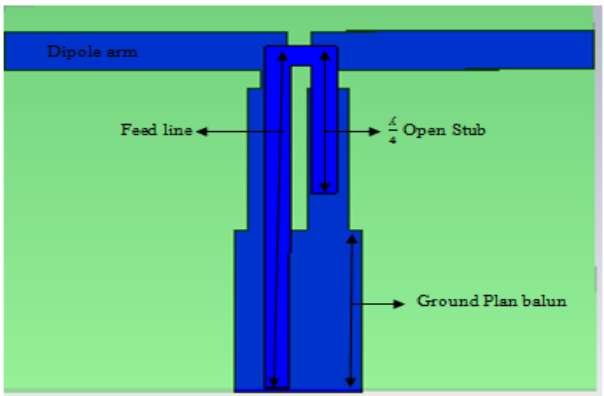

Fig. 1. Dipole with quarter wavelength open stub feed

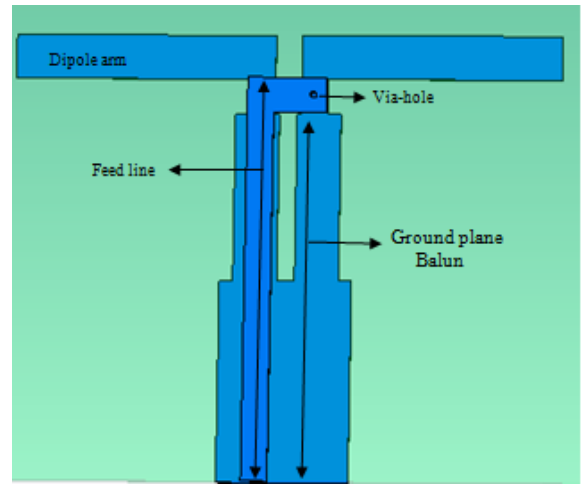

Fig. 2. Dipole with Via-hole feed

TABLE III
DIMENSIONS OF QUARTER WAVLENGTH AND VIA-HOLE

| Parameter | Specifications |
|---|---|
| PCB Substrate | FR-4, h=1.6mm, $\varepsilon_r$ =4.3, tanδ= 0.002. |
| Dipole arm | L= 67mm, W=6mm, gap g = 3mm |
| Microstrip Balun | L= 22mm, L=3mm, w=5mm, w= 3mm |
| Ground Plane | L = 25mm, w= 15mm. |
| Feed Line | L=25mm, w=3mm, Open stub=25mm |
| Dipole arm | Length L= 60mm, Width W=6mm, gap g = 3mm. |
| Via-hole | R= 0.375mm |

III. PRINCIPLES OF BALUN OPERATION AND DESIGN

The main objective of balun design is to provide the system with balance transition between the antenna and its feeding circuit. However, as the rest of the system was then overturned as one part of antenna attached to external plate while the other connects to the internal connector. On the edge attached to the shield, the current can pass through over the outside of the coaxial cable. Baluns changes the tent onto the antenna ports the same magnitude on each but across stage, these tensions caused the same amount of current flow to the outside of the coaxial cable. This type of arrangement will let the feed point to have the same phase as the point of the top shield [6].

Fig. 1 shows quarter wavelength open stub feeds the dipole strips by creating virtual short circuit across the center of the dipole. This arrangement provides further possibilities for reactance compensation of the balance load [7]. The special feature of this method is that it does not require physical direct connection top of substrate. Therefore, it is commonly used to feed microstrip dipole antenna because it requires no soldered or plated through connection [8].

IV. SIMULATION RESULT

A. Quartered wavelength open stub method

The result shows the s-parameter as function of frequency. The plot of S11 shown in Fig. 3 is used to determine whether the antenna is a single band and operating at the desired resonant frequency also the bandwidth can be calculated for the corresponding frequency band. It provides bandwidth of 17% of the resonance frequency with maximum return loss of -47.88 dB as shown in Fig. 3. Fig. 4 illustrates how the resonant frequency is shifted down as the length reduces while Table 4 is a summary of the resulting measurements of different lengths. The balance feeding is illustrated by the value of VSWR in Fig. 5 while Fig. 6 shows the radiation pattern of the antenna.

TABLE IV
COMPARISON FOR DIFFERENT LENGTHS AT W=6 mm

| Length | Z11 | VSWR | RL | BW | Directivity |
|---|---|---|---|---|---|
| 63mm | 45-i 4 | 1.1013 | -26dB | 16% | 1.7dB |
| 65mm | 48-i 2.2 | 1.05 | -32dB | 16.5% | 1.8dB |
| 67mm | 51-i0.216 | 1.0081 | -47.dB | 17% | 2dB |

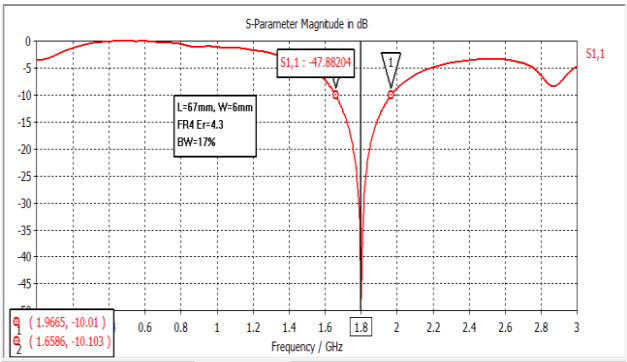

Fig. 3. Return Loss Plot at 1800MHz of antenna

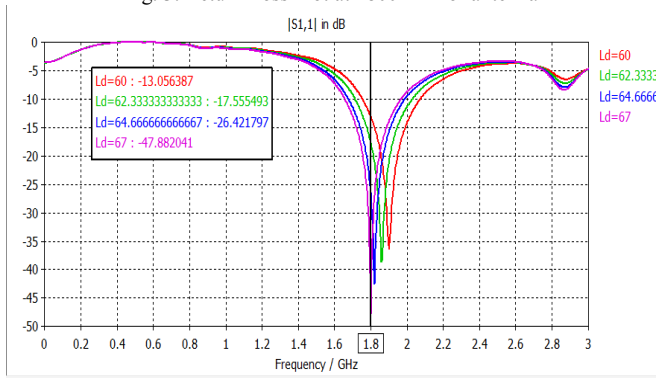

Fig. 4 Return Loss for different lengths of the dipole

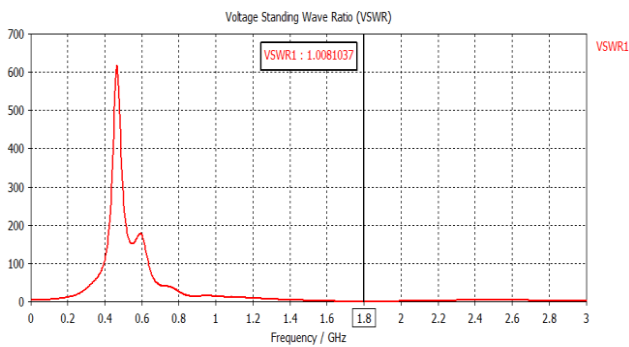

Fig. 5. VSWR of feeding circuit

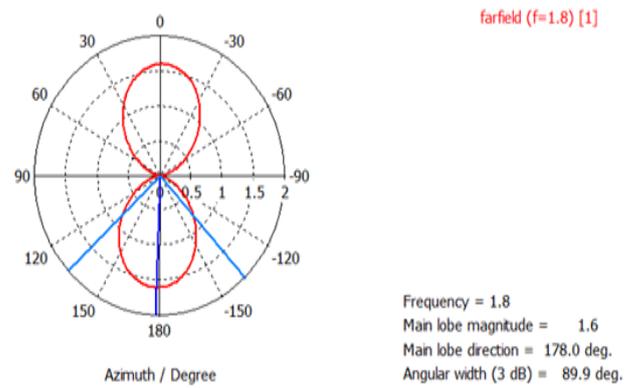

Fig. 6. Directivity in the Azimuth plane

### B. Via-hole feeding method

The dipole radiator is modeled on the top of substrate and adopted with the integrated balun to form a ground plane as a return bath for the current distributions. Since this feeding is through via from the top to the bottom physically connecting the feed line to the center of the dipole to make the formation of center fed the simulations results differs for different widths of the antenna and Table 5 summarize the differences as well as Fig. 8. Besides, the antenna is resonating at 1.8GHz as shown in Fig. 7 with sufficient balance between antenna and its feeding circuit which is determined by the value of VSWR in Fig. 9. From Fig. 10, it can be seen that the antenna has omnidirectional radiation pattern.

TABLE V
COMPARED RESULTS FOR DIFFERENT WIDTHS

| Width | Impedance | VSWR | Return Loss | Bandwidth % |
|---|---|---|---|---|
| 8mm | 43.614 | 1.255 | -18.95dB | 24.5% |
| 7mm | 44.43 | 1.216 | -20.22 dB | 22.5% |
| 6mm | 45.3 | 1.18 | -21.6 dB | 20.6% |
| 5mm | 46.17 | 1.17 | -22.74 | 19.7% |

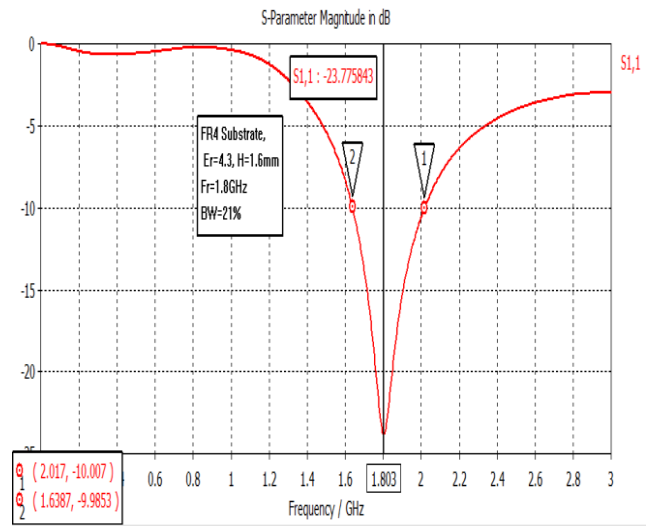

Fig. 7. Return Loss 1800MHz of antenna

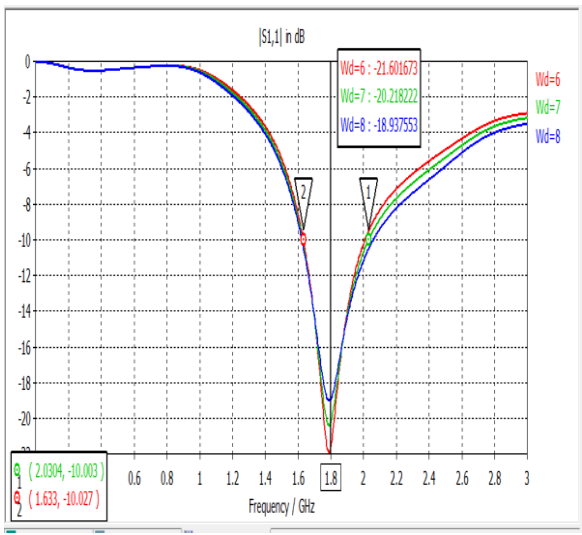

Fig. 8 . Return Loss for different widths of the antenna

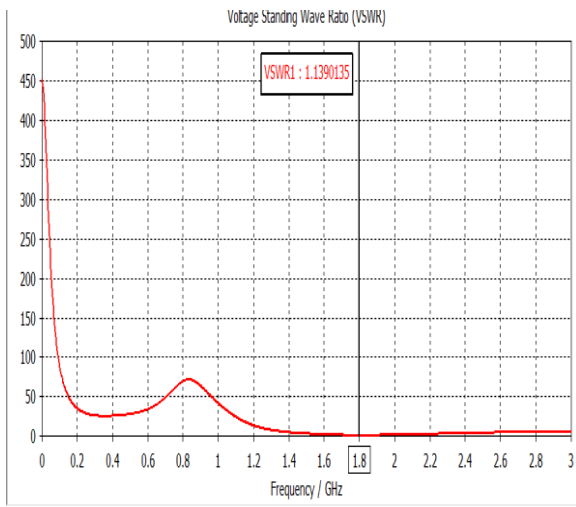

Fig. 9. VSWR of the feed circuit

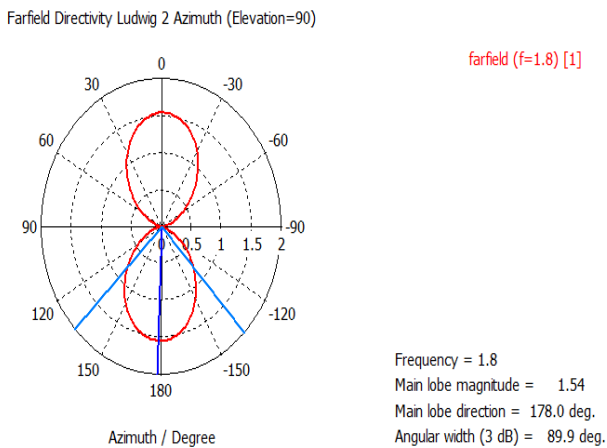

Fig. 10. Directivity of the dipole antenna

Since the polarization of the antenna is horizontal, the E-plane coincides with the azimuth plane and the H-field coincides with elevation plane as shown in Fig. 11 and Fig. 12. The new features in this design in comparison to previous research works is that it has planar structure which can easily integrated to array form to provide greater directivity and high efficiency without causing substantial power losses in the feeding circuit since each element is match to its feeding circuit impedance. In term of bandwidth, this antenna has bandwidth range from 18.4% to 25% while previous search has bandwidth from 11.53% to 13.22%[3]. This advantage is a result of choosing the right configuration of the feeding circuit.

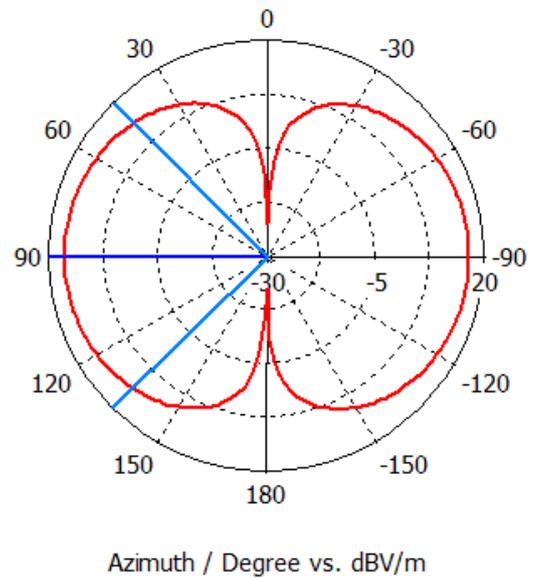

Fig. 11. E-field Radiation at 1.8GHz

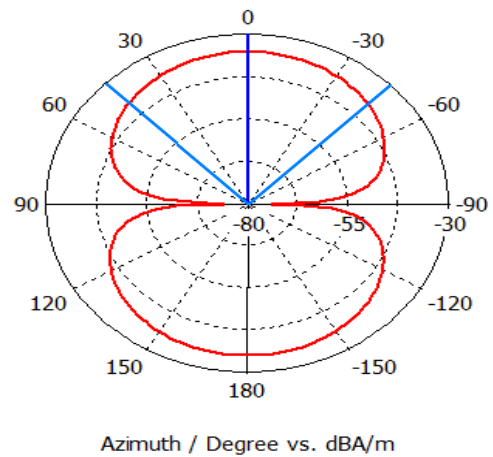

Fig. 12 . H-field Radiation at 1.8GHz

## V. DISCUSSION

Throughout the entire simulation results it has been seen how the antenna performance behaves as function of its geometry parameters. Besides, the feeding technique is important factor that makes substantial effects on the simulation results.

First, the dipole with integrated via-hole balun feed is being used as balance to unbalanced transformer and impedance transformer from coaxial line to the two printed dipoles strip via physical connection. This method is capable to provide us with a bandwidth of 18.4% With maximum return loss of -25dB as shown in Fig. 8 at length =59 mm, width=4 mm and height=1.6 mm. the bandwidth can be increased to 24.5% by changing the width of the antenna from 4mm to 8mm. However, the size will increase and other impacts such as reducing the return loss and increasing the value of VSWR.

In term of practical implementation, it is considered one of the most common suitable due to its compact size and acceptable performance but it requires soldering which may cause junction radiation.

The second part is about open stub integrated balun. This method does not require direct connection between the feed line and the dipole. Therefore, a virtual short circuit is created between the open stub and the dipole strip making a center fed current distribution [7].

The optimized dimension for optimum performance are at length=65 mm, width=6 mm and height=1.6mm. The optimum results are bandwidth of 17 percent with minimum VSWR1.008 v also maximum return loss of -47.88dB. The radiation pattern of this method is unidirectional and has main lobe directivity of 2 dB in the elevation plane with angular width of 89 deg and has the same magnitude on the azimuth plane. For implementation, it is very convenient and suitable since it does not require soldering though it has narrow bandwidth but again it depends on the application requirements.

## V. CONCLUSION

This proposed work focuses on the design and simulation of planar dipole antenna for 1800MHz Band for GSM application using CST studio software. The antenna was structured on FR4 substrate with relative constant of 4.3. Two types of feeding configuration have been designed to feed the antenna in order to match 50 ohm transmission line.

These types are via-hole integrated balun, and quarter wavelength open stub. As results, the dimension of the antenna and simulation results such as return loss, bandwidth and voltage standing wave ratio (VSWR) has been analyzed. For instance, the via-hole is capable to provide maximum return loss of -25dB, bandwidth of 18.4% and VSWR of 1.116 v at length of 59 mm and width 4 mm; it is bandwidth can be improved to 25% and 30%. While the open stub at length of 67mm, width of 6mm and height 1.6mm can provide max return loss of -47.88dB and bandwidth of 17% with VSWR 1.008 << 2 and also has better radiation pattern.

## REFERENCES

[1] Balanis C. A.. 2005.Antenna theory Analysis and design, 3rd edition. Wileterscience publication 3rd. ISBN: 0-471-66782-x.

[2] Garg, R., Bhartia, P., Bahl, I. & Ittipiboon, A. 2001. Microstrip antenna Design handbook. Artech House Boston London. ISBN 0-89006-513-6

[3] Jamaluddin M. H., Rahim, M. K. A., Abd. Aziz, M. Z. A. & Asrokin, A. 2005. Microstrip dipole antenna analysis with different width and length at 2.4 GHz. Asia Pacific conferences on applied electromagnetic, pp. 41-44. DOI: 0-7803-9431-3

[4] Pozar D.M. 2004. Microwave Engineering 3rd edition. , Wiley publication 3rd. ISBN10: 0471448788-

[5] X. Li, Yang, L., Gong, S.X. & Yang, Y.J. 2009. Dual-Band And Wideband Design of A Printed Dipole Antenna Integrated With Dual-Band Balun. *Progress In Electromagnetics Research Letters*, Vol. 6, pp:165-174.

[6] Kuo L.C. , Chuang H.R., Kan Y.-C. & Huang T.-C.2007. A Study of Planar Printed Dipole Antennas For Wireless Communication Application. *J. of Electromagnet. Waves and Appl.,* Vol. , NO 5 .pp: 637-652

[7] Huey-Ru Chuang & Liang-Chen Kuo. 2003. Design Analysis of a 2.4-GHz Polarization-Diversity Printed Dipole Antenna With Integrated Balun and Polarization Switch Circuit For WLAN and Wireless Communication Application. *IEEE Transaction On Microwave Theory and Techniques,*Vol.51(2):374-381, DOI:10.1101/TMT

[8] Shuchita Saxena & Kanaujia, B.K. 2011. Design And Simulation of A Dual Band GapCoupled ANnular Ring Microstrip Antenna, *International Journal of Advances in Engineering & Technology* Vol. 1(2):151-158. ISSN: 2231-1963

[9] Tang, T., , Li, C.M. & Lin C.Y. 2011, Printed Dipole Antenna With back To back Asymmetric Dual-c-Shape Uniform Strip For DTV Application. Progress In Electromagnetics Research C, Vol. 22, pp.73-83.

[10] Oltman G. & Huebner D. A. Member. 1986. Electromagnetically Coupled Microstrip Dipoles, *IEEE Transaction Antennas And Propagation*, Vol. 29(1):151-157. DOI: 0018-926X/81/0100-0151

[11] Shingo Tanaka, Yongho Kim & Hisashi Morishita. 2008. Wideband Planar Folded Dipole Antenna With Self-balance Impedance Property, *IEEE Transaction on Antennas and Propagation*, Vol. 56(5):1223-1228. DOI: 10.1109/TAP